\documentclass[sigconf,screen,nonacm]{acmart}

\setcopyright{none}

\AtBeginDocument{%
  }

\setcopyright{acmlicensed}
\copyrightyear{2027}
\acmYear{2027}
\acmDOI{XXXXXXX.XXXXXXX}
\acmISBN{978-1-4503-XXXX-X/2027/02}
\acmConference[WSDM '27]{The 20th ACM International Conference on Web Search and Data Mining}{February 15--19, 2027}{Hong Kong, China}

\usepackage{booktabs}
\usepackage{multirow}
\usepackage{amsmath}
\usepackage[dvipsnames]{xcolor}
\usepackage{tikz}
\usepackage{adjustbox}
\usetikzlibrary{positioning,arrows.meta,shapes.geometric,fit,calc,backgrounds}
\usepackage{xspace}
\definecolor{ptblue}{HTML}{4477AA}
\definecolor{ptcyan}{HTML}{66CCEE}
\definecolor{ptgreen}{HTML}{228833}
\definecolor{ptyellow}{HTML}{CCBB44}
\definecolor{ptpurple}{HTML}{AA3377}
\definecolor{ptgrey}{HTML}{BBBBBB}

\newcommand{\system}{\textsc{SafeLake}\xspace}
\newcommand{\fullrecall}{\textsc{Full} R@100\xspace}
\newcommand{\productrecall}{Product Recall\xspace}
\newcommand{\tablestyle}{%
  \small
  \setlength{\tabcolsep}{3.5pt}%
  \renewcommand{\arraystretch}{1.08}%
}

\title{Learning the Lake: Reliable Experience for Adaptive Data Product Discovery}

\author{Yixi Zhou}
\authornote{These authors contributed equally to this work.}
\affiliation{%
  \institution{Hong Kong Baptist University}
  \city{Hong Kong SAR}
  \country{China}
}
\email{yxzhou@comp.hkbu.edu.hk}

\author{Fan Zhang}
\authornotemark[1]
\affiliation{%
  \institution{The University of Tokyo}
  \city{Tokyo}
  \country{Japan}
}
\email{zhang-fan@g.ecc.u-tokyo.ac.jp}

\author{Sikun Wang}
\affiliation{%
  \institution{Tokyo University of Science}
  \city{Tokyo}
  \country{Japan}
}
\email{jc19546883@gmail.com}

\author{Yingfan Xu}
\affiliation{%
  \institution{ShanghaiTech University}
  \city{Shanghai}
  \country{China}
}
\email{xuyf2024@shanghaitech.edu.cn}

\author{Haipeng Zhang}
\authornote{Corresponding author.}
\affiliation{%
  \institution{ShanghaiTech University}
  \city{Shanghai}
  \country{China}
}
\email{zhanghp@shanghaitech.edu.cn}

\begin{abstract}
Data-product discovery searches a full lake even when workloads revisit related products and regions. Repetition permits contracted search, but similarity cannot justify a route because one omitted asset invalidates a conjunctive product. We study when serving experience can safely reduce this work. Evolving Discovery Memory records source-labelled query--product--region evidence above a fixed regional index. \system separates operational familiarity, which determines how much to search, from independently calibrated product evidence, which determines where to search. The fixed-probe comparison holds the adaptive budget constant between \system and Familiarity-only. On TAT-QA, product steering raises \productrecall by 0.072; ConvFinQA shows no resolved map gain, while the HybridQA sensitivity favors Familiarity-only in \fullrecall. Trace-only, missing, and false feedback expose boundaries on map steering, while scope-audit agreement cannot certify the source. Across clean confirmed-feedback streams under the frozen transductive protocol, the formal controller saves 49.5--82.7\% of cumulative asset exposure. Experience determines when to contract; reliable evidence determines where to contract.
\end{abstract}


\begin{document}
\maketitle

\section{Introduction}
\label{sec:introduction}

Data-product discovery retrieves the tables and text passages that a downstream application needs for one request. A financial comparison may require a numeric table, a filing passage, and the relation between them. The Data Product Discovery (DPDisc) benchmark formalizes this conjunctive retrieval problem with 13,076 validated instances from HybridQA, TAT-QA, and ConvFinQA~\cite{zhang2026dpdisc}. Its released hybrid retriever starts every request with a full-lake search.

\begin{figure}[t]
  \centering
  \includegraphics[width=\columnwidth]{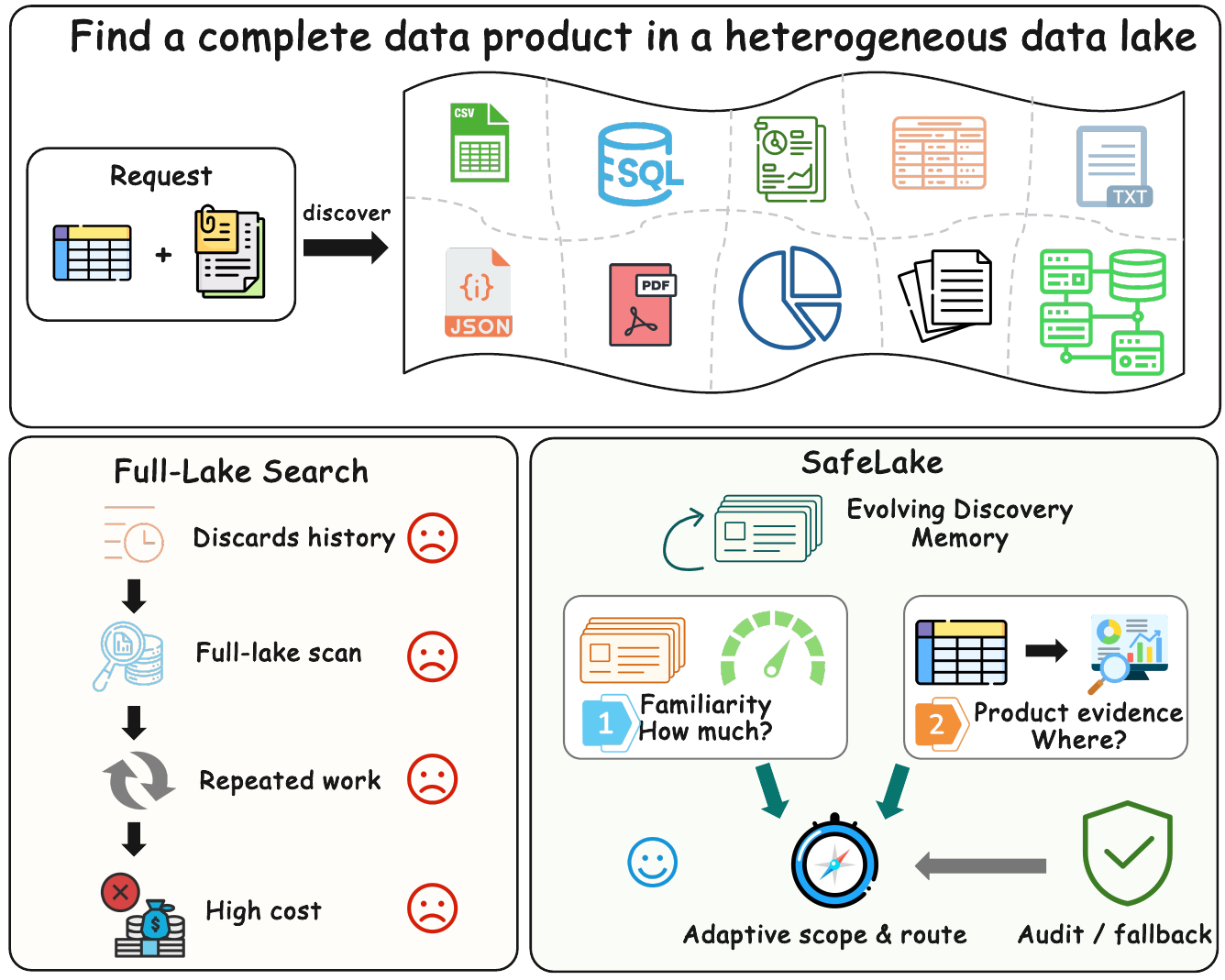}
  \caption{Stateless full-lake retrieval discards serving history and repeats expensive scans. \system uses Evolving Discovery Memory: familiarity selects the search scope, product evidence routes it, and audit or fallback guards the adaptive decision.}
  \Description{A request for a table-and-text data product enters a heterogeneous data lake. The left branch shows Full-Lake Search discarding history, scanning the full lake, repeating work, and incurring high cost. The right branch shows SafeLake using Evolving Discovery Memory. Familiarity controls how much to search, product evidence controls where to search, and audit or fallback guards the adaptive scope and route.}
  \label{fig:problem}
\end{figure}

Serving workloads retain experience that stateless retrieval discards. Analysts revisit companies, periods, and reporting concepts; applications issue families of related requests. Earlier positions in our serve-then-update streams contain 65.9--73.5\% of later gold-asset mentions. With post-hoc product feedback, a service can retain where it found useful products earlier. This observation raises the question that drives our work: \emph{When does reliable experience make later discovery cheaper?}

Figure~\ref{fig:problem} frames this opportunity as control over physical retrieval. Stateless full-lake search discards history and repeats the same scan; a useful memory must instead determine the scope and route while preserving an independent path for audit or fallback. Existing reuse mechanisms cover only part of this combination. Exact replay applies only to identical requests under the same corpus revision. Semantic caches match related requests, but similarity cannot establish multi-product completeness~\cite{dar1996semantic,bang2023gptcache}. Request-local shard and region selection adapts scope from the current query without accumulating source-calibrated evidence across served requests~\cite{aly2013taily,agarwal2026baikal}. In our 72-cell diagnostic, raw-cosine memory saves 98.9\% of asset exposure but loses 0.233 \fullrecall and 0.526 \productrecall. Repetition establishes familiarity; safe routing also requires evidence about useful products.

We call the reusable serving state \emph{Evolving Discovery Memory}. Its query-emergent map records source-labelled query--product--region evidence above a fixed regional index. The regional index is a static physical substrate; the map is workload state that grows during serving. \system separates two decisions: operational familiarity determines \emph{how much} to search, and product evidence from similar requests determines \emph{where} to search. Map steering is conditional: independently calibrated evidence may influence location, while familiarity alone controls only scope. Experience determines when to contract; reliable evidence determines where to contract.

\begin{figure*}[t]
  \centering
  \includegraphics[width=\textwidth]{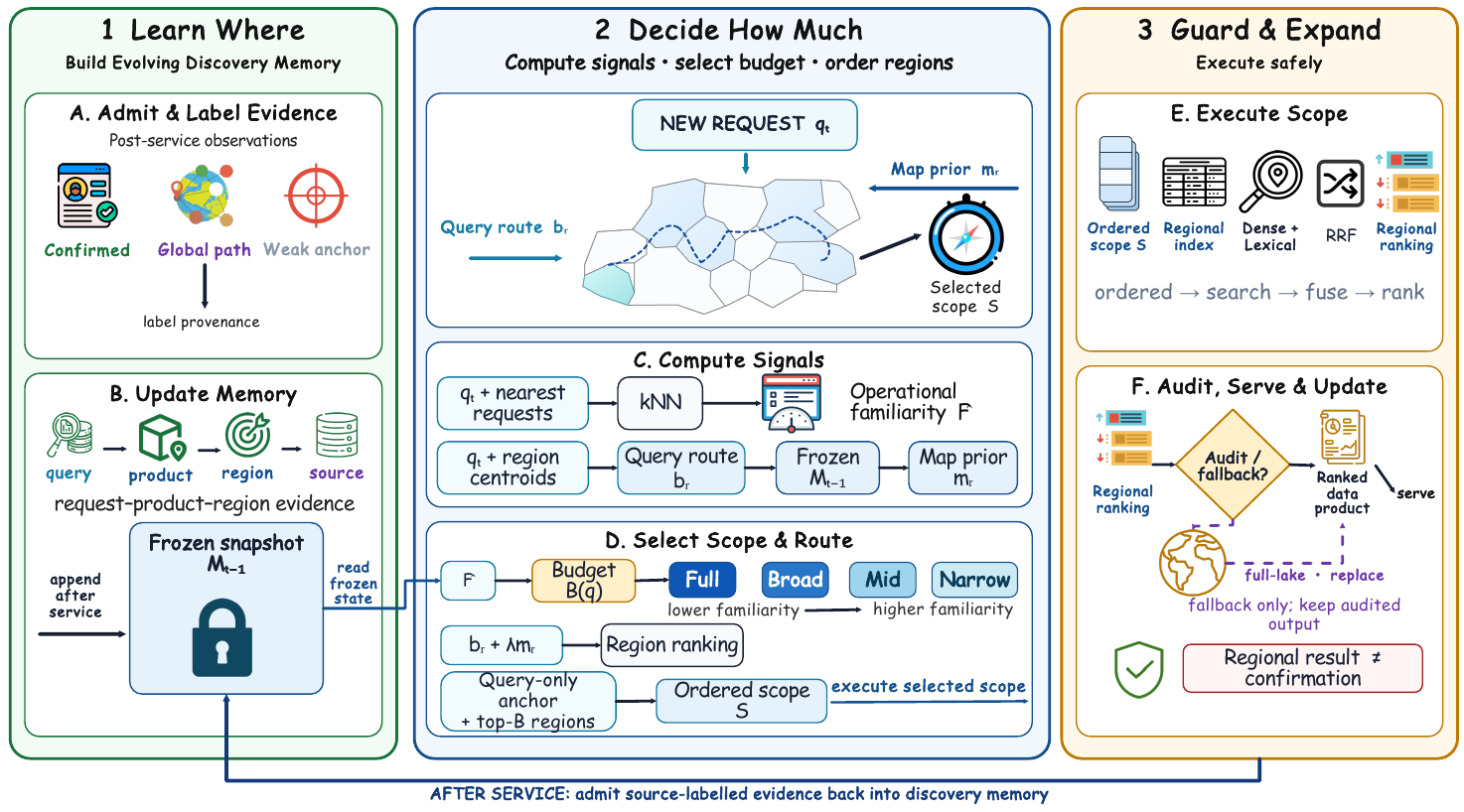}
  \caption{The \system serving pipeline. Learn Where builds frozen, source-labelled memory; Decide How Much selects a budget and ordered regional scope; Guard and Expand executes the selected route, while audit or fallback can return the full-lake result before serving and updating memory.}
  \Description{A three-stage pipeline. Learn Where labels confirmed, global-path, and weak-anchor observations and stores request-product-region-source evidence in frozen memory M sub t minus 1. Decide How Much computes a memory-independent query route from the request and region centroids, and separately reads frozen memory to compute operational familiarity and a map prior. It selects a Full, Broad, Mid, or Narrow budget and builds an ordered scope from a query-only anchor and the top-ranked regions. Guard and Expand executes the regional index using dense and lexical retrieval with reciprocal-rank fusion. Audit or fallback can return the full-lake ranking. The service returns the result before source-labelled evidence enters memory, and a regional result cannot confirm itself.}
  \label{fig:overview}
\end{figure*}

Figure~\ref{fig:overview} makes the serve-then-update boundary explicit. \emph{Learn Where} labels post-service observations as confirmed, global-path, or weak-anchor evidence, then exposes a frozen snapshot $M_{t-1}$ to the next request. \emph{Decide How Much} combines a memory-independent query route with operational familiarity and a map prior; the controller chooses a Full, Broad, Mid, or Narrow budget and orders the resulting scope around a query-only anchor. Within panel C, the two signal computations remain independent: $b_r$ reads only the request and fixed region centroids, whereas $\widehat F$ and $m_r$ read the separate frozen-memory branch. \emph{Guard and Expand} runs the selected scope through the regional index and base retriever. A scheduled audit replaces the regional output with a result from the memory-independent global path (the unchanged full-lake retriever), while fallback routes directly to that path. In panel F, the incoming regional-ranking arrow denotes the audited case; fallback bypasses Execute Scope. The system serves the selected result before it admits new source-labelled evidence, and a regional ranking cannot confirm its own products.

Table~\ref{tab:claim-decomposition} carries our primary mechanism comparison. Familiarity-only retains \system's controller, updates, audits, and every TAT-QA budget decision while removing the map prior. Confirmed product evidence raises \fullrecall by 0.057 and \productrecall by 0.072, with paired intervals that exclude zero. The equal budget isolates where the map routes the contracted scope. ConvFinQA resolves no map gain, and the HybridQA sensitivity favors Familiarity-only in \fullrecall. Boundary studies then test trace-only evidence, missing and false confirmations, audit-gate failures, and workload shift. Under the frozen transductive protocol, clean confirmed-feedback streams save 49.5--82.7\% of cumulative asset exposure. The evidence supports an offline evaluation order: start with static regionalization, add familiarity-based contraction when workloads repeat, and enable product steering only after independent calibration shows a map gain.

This paper makes three contributions:

\begin{itemize}
  \item \textbf{Problem.} We define continual data-product discovery through ordered serve-then-update streams, conjunctive quality, physical work, and cumulative cost. Across three benchmarks, earlier requests cover 65.9--73.5\% of later gold-asset mentions.
  \item \textbf{Principle.} Evolving Discovery Memory separates source-labelled workload state from a fixed regional index. Experience determines when to contract; independently calibrated product evidence determines where to contract.
  \item \textbf{Evidence.} Our primary fixed-probe comparison separates static locality, adaptive budgets, and learned region associations. Boundary diagnostics expose failures under weak, missing, and false evidence; formal-controller streams measure cumulative savings.
\end{itemize}

\section{Related Work}
\label{sec:related}

\paragraph{Data-lake discovery broadens the retrieval target.}
Recent data discovery systems learn semantic representations for unionable or joinable tables. Starmie contextualizes column embeddings for table-union search~\cite{fan2023starmie}; DeepJoin learns representations for equi- and semantic joins~\cite{dong2023deepjoin}; COTER models ad hoc table relevance through conditional optimal transport~\cite{yao2024coter}; and LakeBench evaluates union and join discovery at scale~\cite{deng2024lakebench}. Discovery can feed a second integration stage: ALITE combines tables returned by join, union, and related-table search~\cite{khatiwada2022alite}, while GRAFT jointly retrieves and fuses a connected table subgraph~\cite{ji2026graft}. These systems improve the representation or structure of one request. They treat requests independently and leave cross-request evidence outside their state.

DPDisc moves from one query table to heterogeneous, conjunctive products over tables and text, and supplies our benchmark, metrics, and released retriever~\cite{zhang2026dpdisc}. Baikal allocates a finite budget across semantic lake regions within one complex request~\cite{agarwal2026baikal}. Our question is temporal: after serving one request, can independently sourced product evidence reduce the physical scope of a later request? Request order, update source, conjunctive quality, and cumulative exposure enter the retrieval protocol. This persistent-state perspective follows the broader call to treat memory and execution safeguards as data-system concerns~\cite{ang2026agentic}.

\paragraph{Location evidence differs from cached output.}
Classical semantic data caching stores query-described regions and computes a remainder query for uncovered tuples~\cite{dar1996semantic}. Recent large-language-model (LLM) serving caches operate at different layers: GPTCache reuses a semantically matched response~\cite{bang2023gptcache}, RAGCache retains intermediate states of retrieved knowledge~\cite{jin2025ragcache}, and CacheBlend composes reusable key--value states for retrieved contexts~\cite{yao2025cacheblend}. GroundedCache adds evidence overlap, source-version, and answer-support gates before it reuses a generated answer~\cite{shah2026groundedcache}. These systems cache an answer or model state. Evolving Discovery Memory retains evidence about \emph{where the base retriever should execute}; each non-replay request still runs retrieval over a selected physical scope. Exact replay remains a version-bound special case. A nearby embedding, a reusable hidden state, or a validated answer cannot establish that a fresh multi-product request has all required assets, so our controller preserves an independent global path.

\paragraph{Adaptive retrieval is usually request-local.}
Query-performance prediction estimates the effectiveness of a current request~\cite{he2006query}, while selective search routes that request across shards; Taily models shard score tails for this decision~\cite{aly2013taily}. Baikal extends request-local selection by allocating a finite budget across semantic lake regions~\cite{agarwal2026baikal}. Recent adaptive-retrieval methods condition computation on the current request or generation state. CITADEL routes token interactions through learned lexical keys~\cite{li2023citadel}; Adaptive-RAG selects a strategy from estimated question complexity~\cite{jeong2024adaptive}; Quam expands a reranking pool through query affinity~\cite{rathee2025quam}; and REAPER plans which retrieval sources a complex request should call~\cite{joshi2024reaper}. RetrievalQA and uncertainty-based trigger comparisons show that request-local choices depend on calibration~\cite{zhang2024retrievalqa,moskvoretskii2025adaptive}. These methods adapt inside one request. \system instead tests a longitudinal axis: whether source-labelled evidence accumulated across served requests licenses a smaller physical search. Operational familiarity controls \emph{how much} to search, while calibrated product evidence controls \emph{where} to search.

\paragraph{Serving memory can stay above the index.}
Database cracking reorganizes physical data as a side effect of observed queries. Adaptive metric indexing defers index construction and lets queries refine the structure~\cite{lampropoulos2023adaptive}; CrackIVF applies workload-driven construction to in-memory vector indexes~\cite{mageirakos2025crackivf}; and Quake adapts vector partitions and execution parameters under skew and updates~\cite{mohoney2025quake}. CREAM instead uses soft memory for continual retrieval over a changing document stream~\cite{son2026cream}. These systems adapt the index or retriever. \system fixes the corpus, retriever, and dense-plus-lexical regional index so that the experiment isolates serving experience above the physical substrate. This boundary remains modular: an adaptive index could replace regional execution, and a dynamic-corpus retriever could replace the pinned base model, without changing the source-labelled evidence contract. The distinction lets our evaluation attribute gains separately to static locality, workload familiarity, product feedback, and expansion headroom.

\section{Reliable reuse needs completeness}
\label{sec:problem}

\subsection{Conjunctive discovery needs completeness}

Let a lake contain assets $\mathcal{A}=\mathcal{A}_{t}\cup\mathcal{A}_{x}$, where $\mathcal{A}_{t}$ and $\mathcal{A}_{x}$ denote tables and text. A request $q_i$ has a gold product set $G_i\subseteq\mathcal{A}$. A retriever returns an ordered list $R_i^K$ of at most $K$ assets. We use two complementary metrics:
\begin{align}
  \mathrm{ProductRecall}_i &= \frac{|G_i\cap R_i^K|}{|G_i|}, \\
  \mathrm{FullRecall}_i &= \mathbf{1}[G_i\subseteq R_i^K].
\end{align}
\productrecall measures partial coverage; \fullrecall captures whether the entire data product is available. We follow the pinned DPDisc setup with $K=100$~\cite{zhang2026dpdisc}. The second metric creates the safety requirement: a plausible partial result cannot stand in for a complete result.

The stateless retriever embeds the request, searches dense and sparse indexes for each modality, and fuses rankings with reciprocal-rank fusion~\cite{cormack2009rrf}. We call this path \emph{global} because it exposes the full lake to candidate generation. \system treats its output as the reference path. A full fallback returns the same items in the same order and uses no separately tuned ``full'' mode.

\subsection{Serving history exposes reuse}

Request order determines valid reuse. Given prior requests $q_{<i}$, we define retrospective gold familiarity as
\begin{equation}
  F_{\mathrm{gold}}(i) = \frac{|G_i\cap \bigcup_{j<i}G_j|}{|G_i|}.
\end{equation}
Across all three datasets, earlier requests cover most later gold-asset mentions (Section~\ref{sec:evaluation}). This overlap motivates within-stream reuse. HybridQA retains high train-to-evaluation familiarity, whereas TAT-QA and ConvFinQA use disjoint official asset identifiers across splits. We consequently construct every memory stream in serve-then-update order: request $i$ can use admitted evidence from positions below $i$, never future products. Fixed-probe tests use separate requests with the corpus identity restrictions described in Section~\ref{sec:setup}.

\subsection{Safe reuse needs physical scope}

Three reuse mechanisms occupy distinct points on the frontier. Version-bound exact replay has high precision but few opportunities. A semantic cache matches nearby requests, but its threshold cannot observe a missing product~\cite{dar1996semantic,bang2023gptcache}. A soft map prior can rerank the global result without saving search work. Evolving Discovery Memory targets a narrower physical scope while retaining a path that can contradict its map.

The controller never observes $F_{\mathrm{gold}}$. It computes operational familiarity from earlier request embeddings (Section~\ref{sec:design}). Gold familiarity appears only in this retrospective opportunity analysis.

This observation gives three design requirements. \emph{Where?} Learn Where must label each update by evidence source and use earlier products to rank regions. \emph{How much?} Decide How Much must turn operational familiarity into physical region and asset exposure. \emph{When to trust it?} Guard and Expand must retain query-only evidence, scope audits, and equivalent global fallback, while deployment calibration decides whether a feedback source may steer the map. The system must charge every searched region and asset, including audit work, instead of treating top-$K$ output size as cost.

\section{SafeLake separates map and index}
\label{sec:design}

\system implements the ordered pipeline in Figure~\ref{fig:overview} with two distinct forms of state: a fixed regional index for execution and Evolving Discovery Memory for source-labelled serving evidence. This section defines their interface, the controller that converts evidence into scope, and the independent paths that can override a contracted action.

\paragraph{Two layers.}
The \emph{regional index} is a fixed physical substrate that partitions assets and executes dense and lexical retrieval over a selected region set. The \emph{query-emergent map} is workload state that records source-labelled request--product--region associations. An operator can reset the map without rebuilding the regional index. This distinction separates static locality from experience learned during serving.

\subsection{Serve-then-update blocks self-use}

Every decision for request $q_t$ reads a frozen snapshot $M_{t-1}$. The service creates $M_t$ only after it fixes the returned output and records the source of every new observation. This order prevents the current regional result from influencing its own scope or certifying its own products. It also makes each trace replayable: the snapshot, action, searched assets, returned ranking, and update source fully determine one transition.

The request lifecycle has five steps. First, the stateless query route embeds $q_t$ and scores every fixed region without reading memory. Second, Evolving Discovery Memory finds earlier neighbors and computes operational familiarity and the map prior from $M_{t-1}$. Third, the controller selects a regional budget and builds an ordered scope that retains the query-only anchor. Fourth, Guard and Expand either executes that scope or routes a fallback directly to the global path; a scheduled audit executes both paths and returns the global result. Fifth, the service returns $R_t$ before the ledger admits confirmed, global-path, or weak anchor evidence. A probe stops after the fourth step and leaves the snapshot unchanged.

This lifecycle separates three meanings that a single cache hit would conflate. A request embedding supports familiarity, a product observation supports a region, and an external confirmation supports product utility. The ledger stores these fields separately and assigns each observation an arrival position. Neighbor search can use only positions below $t$. Given the frozen split-level reference mean $\mu$, the controller can reproduce a decision from the trace prefix without future products, updates, or probe labels.

\subsection{Learn Where labels evidence}

Learn Where stores a request embedding, its observed products, their modalities and regions, the request position, and the evidence source. Table~\ref{tab:evidence-contract} fixes the allowed use of each source. \emph{Confirmed evidence} identifies useful products through application feedback. Memory-independent full-lake executions, including fallback and scheduled audits, produce \emph{global-path evidence}. Query-only routes produce \emph{weak anchor evidence} from their selected regions. A path that bypasses memory can still return an incomplete product; routing independence alone cannot certify correctness.

\begin{table}[t]
  \centering
  \caption{Evidence contract. Each source enters after its generating decision. Product utility requires external confirmation.}
  \label{tab:evidence-contract}
  \tablestyle
  \begin{adjustbox}{max width=\columnwidth,center}
    \begin{tabular}{llll}
\toprule
Source & Routing source & Utility label & Allowed map use \\
\midrule
Application feedback & External & Confirmed & Product and scope \\
Global path & Independent & Unconfirmed & Scope check and prior \\
Query-only anchor & Independent & Unconfirmed & Weak prior only \\
\bottomrule
\end{tabular}

  \end{adjustbox}
\end{table}

For a new request $q$, the map retrieves its $k$ nearest historical requests. Neighbor $j$ assigns rank-discounted support
\begin{equation}
  w_j(q)=\frac{\max(0,\cos(e(q),e(q_j)))}{\mathrm{rank}(j)}
\end{equation}
to every product that the observation records. We sum table and text product support within each fixed region and divide the resulting region vector by its largest value to obtain the map prior $m_r(q)$.

The ledger admits every source only after the current request finishes. In deployment, post-hoc confirmation can come from a user selecting or saving an asset, a downstream workflow consuming it, an analyst accepting an assembled product, or curated application feedback. Benchmark gold products instantiate this product-level signal after service. Current-request retrieval never reads them. A source label records provenance; it cannot prove that an application supplied the correct product. A regional output cannot label its own products as confirmed evidence. Probe requests never update the map, so checkpoint changes reflect only the preceding memory stream.

\subsection{Decide How Much turns evidence into scope}

Decide How Much combines a stateless region route with the map prior. Let $b_r(q)=e(q)^\top c_r$ denote cosine similarity between the normalized request and region centroid $c_r$. We min--max normalize $b_r$ across regions and divide $m_r$ by its largest value. The combined score is
\begin{equation}
  s_r(q)=(1-\lambda)\widetilde b_r(q)+\lambda\widetilde m_r(q).
  \label{eq:region-score}
\end{equation}
The final ordered scope first includes a two-region query-only anchor, formed by the two highest $b_r$ values. It then adds regions with nonzero map prior in descending $s_r$ order and fills any remaining slots in descending $b_r$ order. This construction preserves query evidence even when memory assigns a concentrated but incorrect prior.

The map weight $\lambda$ is an admission control. A service uses $\lambda=0$ (Familiarity-only) when it lacks an independent estimate of feedback quality; calibrated sources may use a positive weight. Our fixed-probe study sets $\lambda=0.25$ for the confirmed and trace-only contrasts, then reports where each evidence contract passes or fails. Scope audits measure route agreement; source correctness requires independent calibration.

Let $\mu$ denote the label-blind mean request embedding for the stream split, fixed before serving, and let $\bar e(q)$ be the unit-normalized vector $e(q)-\mu$ (or zero when the difference is zero). For the set $N_8(q)$ of up to eight nearest entries in $M_{t-1}$ under raw cosine similarity, the controller sets $\widehat F(q)=\max_{j\in N_8(q)}\bar e(q)^\top\bar e(q_j)$, with value zero for an empty set, and $n(q)=|N_8(q)|$. It never observes $F_{\mathrm{gold}}(i)$. Let $M$ denote all regions, and let $(b_h,b_m,b_l)$ equal $(8,16,32)$ for HybridQA and TAT-QA or $(16,32,56)$ for ConvFinQA. The formal policy is
\begin{equation}
B(q)=
\begin{cases}
b_h & n(q)\geq3,\ \widehat F(q)\geq0.65,\\
b_m & n(q)\geq3,\ 0.40\leq\widehat F(q)<0.65,\\
b_l & n(q)\geq3,\ 0.20\leq\widehat F(q)<0.40,\\
  M   & \text{else.}
\end{cases}
\label{eq:scope-controller}
\end{equation}
Thus, lower familiarity can only expand scope. This multi-level policy is the formal \system controller evaluated by the fixed-probe study. We retain an earlier one-level regional-or-full controller as a stress-test ablation for broad workload and shift analyses.

The regional budget counts physical regions, while cost counts their assets. This distinction matters because region sizes vary. Every trace records both quantities and charges a regional attempt plus its full audit when both execute.

\subsection{Guard and Expand preserves independence}

Guard and Expand supplies two memory-independent controls. The query-only anchor guarantees that each regional action contains regions that the query route selects without $m_r$. The anchor supplies routing evidence only; completeness requires a scope audit or global fallback. A scheduled \emph{scope audit} executes the unchanged stateless global path, replaces the regional output for that request, and adds global-path evidence for future requests. The audit measures scope agreement with the global path and provides no ground-truth correctness certificate. In particular, agreement cannot detect every false application confirmation. Trace-only service enforces a 5\% minimum cumulative audit rate and therefore cannot reach a zero-global state.

Full fallback returns the same ranked items as the pinned stateless retriever. After serving the request, its output becomes global-path evidence for future requests. The controller routes new or weakly supported regimes through full search until their requests create enough evidence for a regional action. The one-level stress test in Section~\ref{sec:shift} probes whether an existing regional action leaves enough room to expand after a recurring-regime boundary.

Exact replay forms a separate narrow path. Its key hashes normalized request text and the corpus revision; a corpus change invalidates the entry. Semantic neighbors never trigger replay. The evaluation streams contain no eligible exact repeats, so exact replay preserves quality without saving work.

\subsection{Calibration separates scope from steering}

Calibration separates split construction from operating-point selection. We construct each seed-2028 calibration request order and holdout without product labels. After construction, calibration alone reads its labels to score the \fullrecall and \productrecall gates. Static-Calibrated evaluates the dataset-specific budget lists in Section~\ref{sec:setup} and selects the cheapest passing budget. The confirmatory protocol pre-specifies eight neighbors, familiarity thresholds $0.65/0.40/0.20$, the dataset-specific ladders in Equation~\ref{eq:scope-controller}, and map weights $\lambda\in\{0,0.25\}$. Fixed-probe labels evaluate these frozen settings and select none of them.

The two-stage procedure gives $\lambda=0$ a concrete role. It retains the same neighbors, familiarity thresholds, regional budgets, audits, and post-hoc updates as the map-enabled controller, while removing product-based region steering. An operator can therefore fall back to Familiarity-only without discarding accumulated workload history. A deployment may admit a positive map weight only after a paired calibration test preserves the quality gates and improves its chosen quality--cost objective. Source calibration remains external to the query-emergent map because map agreement cannot establish source correctness. Scheduled audits monitor scope coverage and collect memory-independent evidence; source calibration retains promotion authority.

Calibration must follow corpus or retriever revisions. Exact replay already binds entries to a corpus revision. The same revision boundary should trigger a fresh static-budget check and a paired map-weight check because region populations, base rankings, and product evidence may move together. Until those checks pass, the stateless global path and Familiarity-only provide conservative operating points.

\subsection{The regional index executes the scope}

The regional index assigns each table and text asset to one of $M$ embedding clusters. Selecting region set $S$ restricts dense and lexical candidate generation to $\cup_{r\in S}\mathcal{A}_r$, after which the base retriever applies its original reciprocal-rank fusion~\cite{cormack2009rrf}. A static-region baseline ranks regions only by $b_r$ and always uses the same budget. This baseline attributes savings from physical partitioning separately from savings that Evolving Discovery Memory produces.

Corpus updates require a new revision for exact replay and may require reassigning affected assets to regions. Until that rebuild finishes, the controller can route requests through the stateless global path.

\section{Evaluation}
\label{sec:evaluation}

Our evaluation separates confirmatory evidence from boundary diagnostics. The primary line compares the formal multi-level \system controller with Familiarity-only and Static-Calibrated on paired fixed probes; supporting analyses establish the reuse opportunity and measure cumulative cost. The boundary line tests trace-only, missing, and false feedback and action headroom after a workload shift. The 72-cell matrix, one-level controller, and audit gates serve only as diagnostics; the fixed-probe multi-level comparison carries the method claim.

\subsection{Experimental setup}
\label{sec:setup}

\paragraph{Datasets and streams.}
We use the released DPDisc assets and queries from HybridQA, TAT-QA, and ConvFinQA~\cite{zhang2026dpdisc}. The three datasets contain 13,076 validated instances. A broad one-level-controller diagnostic contains 72 cells: three datasets, four query orders, two evidence modes, and seeds 42, 2027, and 2029. \emph{Original} preserves the canonical train order; \emph{shuffled} applies a seeded uniform permutation; \emph{locality-enhanced} concatenates seeded query-embedding cluster blocks; \emph{recurring-to-unseen} serves frequent embedding clusters before disjoint clusters. All constructions are label-blind, and Original represents the natural workload available in DPDisc. Each 400-request stream admits evidence only after serving the current request.

We evaluate the formal multi-level controller with a separate fixed-probe protocol. A seed-2027, 400-request memory stream grows through checkpoints 0, 25, 50, 100, 200, and 400 while the same 300 probes measure each checkpoint without updating the map. HybridQA uses a pre-specified untouched test holdout. TAT-QA and ConvFinQA use deterministic, label-blind train holdouts that exclude the memory stream and normalized duplicate queries. These two holdouts diagnose the mechanism and carry no untouched-test claim.

The two evidence modes have different contracts. Confirmed feedback uses benchmark gold products after service to simulate the post-hoc usage signals described in Section~\ref{sec:design}. Current-request retrieval never sees these products. Trace-only memory receives query-only anchor outputs plus global-path evidence from scope audits. Probe gold serves evaluation only. In each nested missing- or false-feedback schedule, every higher-rate selected set contains all lower-rate events. False-confirmation controls replace 0\%, 5\%, 10\%, or 20\% of events with a product from another stream request after removing every asset shared with the current gold product. The two pre-specified randomization schedules share seed 2027, fixed probes, the backend, and a 5\% target audit rate, but use distinct namespaces. At stream position $j$, a SHA-256 draw over the seed, namespace, position, and request identifier selects an audit; a cumulative floor enforces at least $\lfloor0.05j\rfloor$ audits.

\paragraph{Retriever and configurations.}
The pinned base retriever combines Granite embeddings, lexical retrieval, and reciprocal-rank fusion. We pin DPDisc commit \texttt{c4b466a}, dataset revision \texttt{aef6ff4}, and Granite revision \texttt{4ab61ff}. Its \fullrecall differs from the DPDisc paper by 0.0246, 0.0390, and 0.0692 on HybridQA, TAT-QA, and ConvFinQA. Our artifact audit byte-matches the corpus, ground truth, and evaluator; a modality check reproduces reported TAT-QA text recall but localizes the remaining gap to the released table retrieval/index path. ConvFinQA also requires batch 256 because the released batch 512 requires more than 24\,GiB. We cannot isolate the remaining published-to-code delta, so every comparison uses a stateless result from the same backend.

The new claim-focused controls use a deterministic exact reciprocal-rank-fusion reconstruction (Exact-RRF): exhaustive Granite inner-product and lexical rankings feed the same fusion. We keep every comparison within this backend. Its stateless \fullrecall result agrees with the retained aggregate within one probe on TAT-QA and ConvFinQA and falls 0.010 lower on HybridQA. We mark the HybridQA rows as sensitivity results.

The formal controller uses 256, 64, and 128 regions, eight memory neighbors, and $\lambda=0.25$ in Equation~\ref{eq:region-score}. Its monotonic ladders are 8/16/32/full for HybridQA and TAT-QA and 16/32/56/full for ConvFinQA. We freeze the thresholds and ladders before the fixed-probe comparison. The earlier one-level 32/32/56-or-full controller remains a stress-test ablation; the multi-level controller carries the method claim.

\paragraph{Calibration and freezing.}
Request ordering and holdout construction use no product labels. After construction, a separate seed-2028 calibration set per dataset uses its labels only to score the \fullrecall and \productrecall gates. Static-Calibrated scans $8/16/32/64/128/256$ regions on HybridQA, $8/16/32/64$ on TAT-QA, and $8/16/32/56/64/96/128$ on ConvFinQA; the last value in each list is full search. It selects the cheapest passing budget. The formal comparison freezes eight neighbors, thresholds $0.65/0.40/0.20$, both dataset-specific ladders, audit schedules, and $\lambda\in\{0,0.25\}$. Fixed-probe and test labels evaluate these settings and select none of them.

\paragraph{Primary controls and metrics.}
The primary fixed-probe comparison uses stateless full, \system, Familiarity-only, and Static-Calibrated. \emph{Familiarity-only} retains the formal ladder, thresholds, updates, and audits while ranking regions only by the query route. \emph{Static-Calibrated} freezes the budget selected on seed-2028 calibration requests. \emph{Static-Matched} chooses the closest evaluation cost and supports descriptive frontier analysis only. \fullrecall measures conjunctive completeness; \productrecall measures partial product coverage. Search ratio divides physically exposed assets, including scope audits, by the full lake size. The three independent terminal criteria require \fullrecall drop $\leq0.02$, \productrecall drop $\leq0.01$, and search ratio $\leq0.70$. The implementation also records the equivalent global-saving check $1-\text{search ratio}\geq30\%$.

\paragraph{Matched controls isolate three mechanisms.}
The comparison changes one decision at a time. Map-prior full changes region scores but exposes the entire lake, so any quality change comes from reranking and cannot count as physical saving. Static-Calibrated fixes a regional budget before evaluation, which isolates the saving available from the regional index without per-request familiarity. Familiarity-only shares every formal-controller action with \system except the product-derived map prior, so its paired difference isolates where reliable product evidence routes a fixed scope. Static-Matched answers a different descriptive question by locating the nearest observed cost on the full static frontier. We keep it outside the pre-specified claim because evaluation requests choose its budget.

\paragraph{Boundary diagnostics.}
Version-bound exact replay, raw-cosine memory, calibrated semantic cache, map-prior full, periodic fallback, and the one-level controller test output reuse, reranking, audit cost, and action headroom. Raw-cosine memory and semantic caching reuse an earlier output when query embeddings cross a threshold; exact replay additionally requires identical normalized text and corpus revision. These methods may save nearly all retrieval work because they bypass fresh candidate generation, although they cannot observe an omitted member of a conjunctive product. We report their \fullrecall and \productrecall without applying the regional-controller gates to their internal scores. These boundary methods carry no formal quality claim; the fixed-probe controls carry the mechanism claim, and the clean controller streams carry the cost claim.

The audit-gate diagnostic measures $d=C(S_{\mathrm{map}})-C(S_{\mathrm{query}})$ after service, where $C(S)$ is the fraction of top-100-per-modality global-path outputs covered by scope $S$. The rolling gate requires trusted neighbors, at least three audits, positive mean $d$ over the latest ten, and more positive than negative residuals. After its recorded failure, we froze a cumulative gate with a 400-request window and 0.01 margin. Each gate uses its frozen promotion rule and remains outside the primary comparison.

\paragraph{Execution and integrity.}
Every run records its configuration, code hash, stream identifiers, per-request actions, searched regions and assets, evidence source, scope audit, latency, quality, update time, and memory size. The paper-facing aggregates and their provenance hashes are frozen for artifact release. A fresh execution passes all 42 invariant tests, covering corpus-bound replay, fallback equivalence, paired audit schedules, nested missing and false confirmations, minimum audit rate, and rejection of self-confirmed updates.

The fixed-probe protocol reuses the same probes across checkpoints and paired methods. Query-paired bootstrap intervals therefore resample probes, preserve within-query differences, and condition on the frozen stream, controller, and regional index. The 72-cell diagnostic varies seeds and label-blind request orders to expose workload sensitivity; it supports boundary analysis instead of the formal method comparison. We state a resolved paired effect only when its confidence interval excludes zero, and we retain gate failures even when another metric improves.

\subsection{Serving history creates reuse opportunities}

Earlier requests cover 73.45\% of later gold-asset mentions on HybridQA, 65.93\% on TAT-QA, and 71.79\% on ConvFinQA. Prior product-overlap rates reach 77.06\%, 69.51\%, and 75.15\%. These values measure opportunity under serve-then-update order: request $i$ observes positions below $i$ and never future products. Their scope remains within-stream; cross-split transfer remains untested. TAT-QA and ConvFinQA use disjoint official asset identifiers across splits, which motivates the within-stream and fixed-probe protocols.

\subsection{Reliable evidence improves TAT-QA routing}

\begin{table}[t]
  \centering
  \caption{Supporting boundary diagnostic averaged over the 72-cell one-level matrix. Quality differences are method minus stateless; positive values favor the method. Asset saving includes every executed scope.}
  \label{tab:method-frontier}
  \tablestyle
  \begin{adjustbox}{max width=\columnwidth,center}
    \begin{tabular}{lrrr}
\toprule
Method & $\Delta$ Full R@100 & $\Delta$ Product Recall & Assets saved \\
\midrule
Exact replay & 0.0000 & 0.0000 & 0.0\% \\
Raw-cosine memory & $-$0.2331 & $-$0.5256 & 98.9\% \\
Calibrated similarity cache & $-$0.0272 & $-$0.1058 & 60.3\% \\
Map-prior full & +0.0140 & +0.0060 & 0.0\% \\
Static region & +0.0047 & $-$0.0020 & 56.2\% \\
Periodic fallback & +0.0044 & $-$0.0020 & 53.4\% \\
One-level controller & +0.0050 & $-$0.0010 & 50.5\% \\
\bottomrule
\end{tabular}

  \end{adjustbox}
\end{table}

Table~\ref{tab:method-frontier} supplies a supporting boundary diagnostic for output reuse, global reranking, and physical scope. Exact replay finds no eligible repeat. Raw-cosine memory saves nearly all work but loses conjunctive coverage. A map prior over full retrieval preserves quality without reducing scope. Static regions provide the main physical saving. The one-level controller searches more than its static action because it charges audits; this table excludes formal-method comparison.

Static-Calibrated freezes the calibration-selected budgets at 8 for HybridQA, 32 for TAT-QA, and 56 for ConvFinQA. HybridQA budget 8 then incurs a 0.014 \productrecall drop on fixed probes; we retain this miss without retuning. Static-Matched selects the closest evaluation cost and supports descriptive frontier analysis only.

\begin{table*}[t]
  \centering
  \caption{Primary formal comparison: confirmed \system minus two controls on fixed probes. Familiarity-only uses the same adaptive budget and paired audits; Static-Calibrated freezes its budget after calibration. Negative search differences favor \system. Query-paired bootstrap uses 10,000 samples for 95\% confidence intervals (CIs). $^\dagger$HybridQA uses the claim-focused reconstruction whose stateless \fullrecall differs from the retained endpoint by 0.010; its rows are sensitivity results.}
  \label{tab:claim-decomposition}
  \tablestyle
  \begin{adjustbox}{max width=\textwidth,center}
    \begin{tabular}{llrrr}
\toprule
Dataset & Baseline & $\Delta$ Full R@100 [95\% CI] & $\Delta$ Product Recall [95\% CI] & $\Delta$ search [95\% CI] \\
\midrule
\multirow{2}{*}{HybridQA$^\dagger$}
 & Familiarity-only & $-.013\ [-.027,-.003]$ & $-.005\ [-.014,+.004]$ & $-.009\ [-.010,-.007]$ \\
 & Static-Calibrated & $+.003\ [-.017,+.023]$ & $+.017\ [+.004,+.031]$ & $+.105\ [+.078,+.134]$ \\
\multirow{2}{*}{TAT-QA}
 & Familiarity-only & $+.057\ [+.023,+.093]$ & $+.072\ [+.047,+.098]$ & $+.005\ [+.002,+.007]$ \\
 & Static-Calibrated & $+.053\ [+.023,+.087]$ & $+.016\ [-.001,+.032]$ & $-.352\ [-.380,-.321]$ \\
\multirow{2}{*}{ConvFinQA}
 & Familiarity-only & $-.007\ [-.020,+.007]$ & $+.000\ [-.009,+.008]$ & $-.006\ [-.008,-.004]$ \\
 & Static-Calibrated & $-.003\ [-.013,+.007]$ & $-.002\ [-.012,+.005]$ & $-.276\ [-.304,-.246]$ \\
\bottomrule
\end{tabular}

  \end{adjustbox}
\end{table*}

Table~\ref{tab:claim-decomposition} is the primary formal comparison and separates adaptive budget selection from learned region associations. SafeLake and Familiarity-only share every TAT-QA budget decision, with a mean of 10.24 regions. Confirmed product evidence lets the map raise \fullrecall by 0.057 and \productrecall by 0.072; both intervals exclude zero. ConvFinQA shows no resolved quality difference, and the HybridQA sensitivity favors Familiarity-only in \fullrecall. The comparison supports conditional map authority: reliable evidence improves where a fixed scope goes on confirmed TAT-QA.

Static-Calibrated supplies the pre-specified comparison. On TAT-QA, \system raises \fullrecall by 0.053 and searches 35.2 percentage points fewer assets. On ConvFinQA, both quality intervals include zero while search falls by 27.6 points. The earlier Static-Matched frontier remains descriptive: its clearest point is TAT-QA, where confirmed \system changes \productrecall by $+0.042$ and search by $-0.061$.

The one-level ablation supplies a final diagnostic. Across its broad stress test, all 24 dataset--workload--evidence aggregates pass the three independent gates; 65 of 72 individual cells pass. The seven \productrecall failures on TAT-QA or ConvFinQA remain visible because \fullrecall can hide one omitted asset from a larger product. This ablation cannot assign smaller scopes to more familiar requests. The fixed-probe study carries the formal method comparison.

\subsection{Feedback reliability bounds map authority}
\label{sec:evolution}

\begin{table*}[t]
  \centering
  \caption{Fixed-probe evolution from checkpoint 0 to 400. Gold prior is an offline diagnostic that measures map mass on gold-asset regions; the controller never observes it. Drops equal stateless minus \system, so positive values indicate loss.}
  \label{tab:evolution}
  \tablestyle
  \begin{adjustbox}{max width=\textwidth,center}
    \begin{tabular}{llrrrrr}
\toprule
Dataset & Evidence & Familiarity & Gold prior & Budget & Search ratio & Full R@100/Product Recall drop \\
\midrule
\multirow{2}{*}{HybridQA}
 & Confirmed & .578 & .372 & $256\!\to\!14.8$ & $1.00\!\to\!.134$ & $-.017/-.009$ \\
 & Trace-only & .578 & .341 & $256\!\to\!14.8$ & $1.00\!\to\!.133$ & $-.003/-.001$ \\
\multirow{2}{*}{TAT-QA}
 & Confirmed & .744 & .601 & $64\!\to\!10.2$ & $1.00\!\to\!.246$ & $-.043/+.009$ \\
 & Trace-only & .744 & .387 & $64\!\to\!10.2$ & $1.00\!\to\!.250$ & $+.030/+.069$ \\
\multirow{2}{*}{ConvFinQA}
 & Confirmed & .723 & .446 & $128\!\to\!22.1$ & $1.00\!\to\!.305$ & $-.010/-.004$ \\
 & Trace-only & .723 & .308 & $128\!\to\!22.1$ & $1.00\!\to\!.306$ & $.000/+.007$ \\
\bottomrule
\end{tabular}

  \end{adjustbox}
\end{table*}

Table~\ref{tab:evolution} shows that all six settings accumulate similar operational familiarity, while their product evidence differs. Familiarity reaches 0.578--0.744, and mean budget falls by 83--94\%. Confirmed evidence assigns 0.372, 0.601, and 0.446 prior mass to gold regions at checkpoint 400. The trace-only map assigns less mass on every dataset.

Five settings preserve the quality tolerances while their search ratio falls to 0.133--0.306. TAT-QA trace-only fails: \fullrecall drop reaches 0.030 and \productrecall drop reaches 0.069. Its budget still contracts from 64 to 10.2 regions. Query similarity grows while weak anchors dominate the map, producing high operational familiarity without reliable conjunctive coverage.

Confirmed TAT-QA illustrates the opposite case. It searches 24.6\% of assets at checkpoint 400, improves \fullrecall by 0.043, and keeps \productrecall drop at 0.009. The same requests and ladder produce the trace-only failure; the update contract is the only changed condition.

\begin{table}[t]
  \centering
  \caption{TAT-QA feedback reliability at checkpoint 400. Retention rows use one nested schedule; false-feedback rows give the range and passing schedules across two nested schedules. Familiarity is .744, mean base budget is 10.24 regions, and search ratio is .260--.278.}
  \label{tab:partial-confirmation}
  \tablestyle
  \begin{adjustbox}{max width=\columnwidth,center}
    \begin{tabular}{llrrl}
\toprule
Perturbation & Target & Realized & Product Recall drop & Gate \\
\midrule
Retain true feedback & 0\%   & 0.0\%  & .048 & fail \\
 & 25\%  & 23.3\% & .048 & fail \\
 & 50\%  & 47.5\% & .036 & fail \\
 & 75\%  & 74.0\% & .010 & pass \\
 & 100\% & 100\%  & $-.003$ & pass \\
\midrule
Inject false feedback & 0\%  & 0.0\% & $[-.003,-.003]$ & 2/2 \\
 & 5\%  & 4.5--6.5\% & $[-.002,.003]$ & 2/2 \\
 & 10\% & 8.3--11.5\% & $[-.001,.006]$ & 2/2 \\
 & 20\% & 22.3--23.0\% & $[.009,.011]$ & 1/2 \\
\bottomrule
\end{tabular}

  \end{adjustbox}
\end{table}

Table~\ref{tab:partial-confirmation} holds the familiarity ladder fixed and changes feedback reliability. Retaining 0\%, 25\%, or 50\% of true confirmations misses the \productrecall gate; 75\% and 100\% pass. \productrecall drop falls from 0.048 to 0.010 at 75\%. False confirmations test a different failure: each corrupted event inserts a disjoint wrong product. Both schedules pass through a 10\% target rate. At 20\%, \productrecall drop reaches 0.009--0.011 and one schedule fails. ConvFinQA passes both schedules at every rate; the HybridQA sensitivity, where the map has no resolved clean advantage, becomes schedule-sensitive at 5\% and fails both schedules at 10\% and 20\%.

Audit agreement yields unstable branch selection. The rolling-gate policy meets the three independent terminal criteria in 9 of 12 dataset--noise cells; the cumulative-gate policy meets them in 10 of 12. The first keeps the TAT-QA map active at 20\% corruption, while the second rejects it even with clean feedback and incurs a 0.065 \productrecall drop at the small familiarity budget. Neither gate meets its promotion rule. We therefore retain source calibration as the admission control for $\lambda$ and report automatic switching as a negative diagnostic.

Shuffling the learned request--product associations provides a direct map ablation on HybridQA. At checkpoint 400, gold-region prior mass falls from 0.372 to 0.016 with confirmed evidence and from 0.341 to 0.015 in trace-only mode. This collapse shows that the map's region mass comes from learned associations instead of marginal region popularity.

\subsection{Action headroom bounds recovery}
\label{sec:shift}

This boundary experiment uses the one-level controller as a stress-test instrument. We restrict its interpretation to action headroom and exclude it from the main method comparison. A label-blind partition serves 200 recurring requests from regime A, then 200 from regime B. On HybridQA, familiarity falls from 0.745 to 0.474; search ratio rises from 0.216 to 0.268 and returns to 0.207 at steady B. All three seeds reach the recovery criterion after 10--13 requests.

TAT-QA and ConvFinQA begin with actions that already expose 62--63\% of assets. Their immediate search ratio increases by 0.030 and 0.013, followed by inconsistent contraction. The pre-specified HybridQA stop-loss required a 0.10 increase, while the observed increase is 0.052. We retain this failed threshold. In this one-level stress test, observable recovery appears only where the action has room to expand.

\subsection{Formal control amortizes by request six}

\begin{table}[t]
  \centering
  \caption{Formal cumulative cost and overhead on clean confirmed-feedback streams under two pre-specified 5\% audit schedules. Values common after rounding appear once; ranges give their minimum and maximum. Break-even assumes an existing regional index.}
  \label{tab:cumulative}
  \tablestyle
  \begin{adjustbox}{max width=\columnwidth,center}
    \begin{tabular}{lrrrrr}
\toprule
& \multicolumn{2}{c}{Confirmed stream} & \multicolumn{3}{c}{Formal control} \\
\cmidrule(lr){2-3}\cmidrule(lr){4-6}
Dataset & Break-even & Saving & Control & Update & Memory \\
 & (queries) & (400 queries) & p95 (ms) & p95 (ms) & (MiB) \\
\midrule
HybridQA & 6 & 82.7\% & 5.86 & .067 & 1.83 \\
TAT-QA & 4 & 61.6\% & 2.38 & .036 & 1.36 \\
ConvFinQA & 4 & 49.5--50.5\% & 3.14 & .040 & 1.37 \\
\bottomrule
\end{tabular}

  \end{adjustbox}
\end{table}

Formal-controller accounting includes cold full requests, regional attempts, and scope audits. On clean confirmed-feedback streams with an existing regional index, cumulative asset exposure crosses below repeated stateless search after four requests on TAT-QA and ConvFinQA and six on HybridQA (Table~\ref{tab:cumulative}). Over 400 requests, the formal controller saves 82.7\% on HybridQA, 61.6\% on TAT-QA, and 49.5--50.5\% on ConvFinQA across the two audit schedules.

At checkpoint 400, controller p95 ranges from 2.38 to 5.86 ms and confirmed updates remain below 0.067 ms. Confirmed state occupies 1.36--1.83 MiB; weak-anchor state occupies 4.53--5.05 MiB because it stores longer retrieved lists. The microbenchmark includes neighbor lookup, familiarity, budget selection, map prior construction, and regional route construction.

\begin{table}[t]
  \centering
  \caption{Isolated regional-retrieval latency over 200 requests per dataset, excluding control, updates, and audits.}
  \label{tab:latency}
  \tablestyle
  \begin{adjustbox}{max width=\columnwidth,center}
    \begin{tabular}{lrrr}
\toprule
Dataset & p95 reduction & Search ratio & Region budget \\
\midrule
HybridQA & 81.0\% & 15.9\% & 32 \\
TAT-QA & 28.6\% & 58.1\% & 32 \\
ConvFinQA & 32.9\% & 57.9\% & 56 \\
\bottomrule
\end{tabular}

  \end{adjustbox}
\end{table}

The isolated paired runs connect asset scope to wall-clock retrieval. HybridQA searches 15.9\% of assets and reduces p95 latency by 81.0\%. TAT-QA and ConvFinQA search about 58\% and reduce p95 by 28.6\% and 32.9\%. These runs alternate action order and execute one method at a time to avoid concurrency and cache-order artifacts. They characterize the regional substrate and exclude controller, update, and scope-audit latency.

\section{Discussion}
\label{sec:discussion}

\paragraph{Evidence quality controls safe contraction.}
The confirmation and corruption tests identify only dataset-specific operating boundaries. Their nonmonotonic response, together with failed audit-residual promotion gates, shows that route agreement cannot infer feedback truth. A deployment should therefore estimate application feedback quality outside the map, retain $\lambda=0$ when that estimate is unavailable, and pair weaker sources with broader budget floors or more frequent audits.

\paragraph{The three sources of saving remain separable.}
Static regionalization supplies physical locality, operational familiarity adapts the budget, and product evidence reranks regions within that budget. Only confirmed TAT-QA resolves a benefit from the third mechanism; ConvFinQA and the HybridQA sensitivity do not. Familiarity-only is therefore a useful operating point that retains adaptive scope and accumulated workload history when paired calibration finds no map benefit.

\paragraph{Recovery requires action headroom.}
The HybridQA stress test exposes an expand--learn--contract pattern because its one-level controller leaves enough range between regional and full actions. TAT-QA and ConvFinQA begin from substantially broader regional actions, which suppresses measurable expansion, and the HybridQA increase misses the pre-specified amplitude gate. The result diagnoses the tested action ladder and provides no general recovery guarantee for the formal controller.

\section{Limitations}
\label{sec:limitations}
Our evidence covers three question-answering-derived DPDisc lakes with fixed membership and one pinned retriever, so corpus updates, other retrieval stacks, and cross-lake transfer require recalibration. TAT-QA and ConvFinQA use label-blind within-train holdouts, whereas HybridQA alone supplies an untouched-test mechanism result. Centered familiarity uses the complete split-level query-embedding mean as a label-blind transductive reference; a deployment must estimate this mean from an earlier workload. Asset exposure approximates work, and the latency study excludes end-to-end serving. The feedback controls omit delayed, correlated, and population-dependent errors; global-route audits cannot certify completeness. Paired intervals condition on frozen streams, probes, and budgets: the controller guarantees monotonic scope selection without guaranteeing recall preservation or shift detection.

\section{Conclusion}

Under our frozen transductive protocol, repeated data-product discovery reuses serving experience by separating familiarity-based scope from calibrated product steering. Fixed probes resolve steering value only on confirmed TAT-QA. Deployment therefore requires paired calibration and a centering mean estimated from earlier workload data.

\clearpage
\section*{Ethical Considerations}

Omissions can affect financial or enterprise decisions. Quality gates, audits, replay, and fallback reduce risk without guaranteeing completeness: provenance cannot verify feedback, and global retrieval may miss assets. Operators should set harm-based targets and audit rates, use Familiarity-only without calibrated map value, retain fallbacks, and require human review. Workload memory can disclose sensitive queries and proprietary links. Deployments should minimize retained fields, separate logs from embeddings, enforce tenant-aware access and deletion, and record provenance. We use public benchmarks without human subjects; retrieval evaluation is not financial advice. We used a large language model (LLM) only to polish the writing of this paper.

\bibliographystyle{ACM-Reference-Format}
\bibliography{references}

\end{document}